\documentclass[aps,pra,final,10pt,twocolumn,superscriptaddress]{revtex4-1}
\usepackage{amsmath,amssymb,bm}
\usepackage{graphicx,color}
\usepackage[squaren]{SIunits}
\usepackage[english]{babel}
\usepackage{amstext}
\usepackage{amsthm}
\usepackage{latexsym}
\usepackage{array}
\usepackage{color}
\usepackage{float}
\usepackage{microtype}
 
\usepackage{multirow}
\usepackage{dcolumn}
\definecolor{url}{RGB}{0,20,160}
\usepackage[colorlinks=true,linkcolor=blue,citecolor=blue,urlcolor=url]{hyperref}
\usepackage[usenames,dvipsnames,svgnaes,table]{xcolor}
\usepackage{graphicx, type1cm, lettrine}
\usepackage[utf8]{inputenc}

\usepackage{natbib}
\makeatletter
\def\frutiger{cmss10 }
\def\frutigerbold{cmssbx10 }
=\frutigerbold at 14pt
=\frutiger at 8pt
=\frutigerbold at 12pt
=\frutigerbold at10pt
=\frutigerbold at 8pt
=\frutiger at 8pt
=\frutigerbold at 31pt
\def\@caption@tabnum@sep{\figtextfont{{ }{\bf\textbar}{ }}}%
\def\fnum@table{{\bf\tablename~\thetable}}

\def\@caption@fignum@sep{\figtextfont{{ }{\bf\textbar}{ }}}%
\def\fnum@figure{{\bf\figurename~\thefigure}}
\renewenvironment{figure}{\@float{figure}\def\textbf##1{{\fignumfont ##1}}\def\bf{\fignumfont}}{\end@float}
\def\@startsection#1#2#3#4#5#6{%
	\if@noskipsec\leavevmode\fi
	\par\@tempskipa #4\relax
	\@afterindenttrue
	\ifdim\@tempskipa <\z@
	\@tempskipa -\@tempskipa \@afterindentfalse
	\fi\if@nobreak\everypar{}%
	\else\addpenalty\@secpenalty\addvspace\@tempskipa\fi
	\@ifstar{\@ssect{#3}{#4}{#5}{#6}}{\@dblarg{\@sect{#1}{#2}{#3}{#4}{#5}{#6}}}}
\def\@sect#1#2#3#4#5#6[#7]#8{%
	\ifnum #2>0
	\let\@svsec\@empty
	\else\refstepcounter{#1}\protected@edef\@svsec{\@seccntformat{#1}\relax}\fi
	\@tempskipa #5\relax
	\ifdim\@tempskipa>\z@
	\begingroup#6{\@hangfrom{\hskip #3\relax\@svsec}%
		\interlinepenalty \@M #8\@@par}\endgroup
	\csname #1mark\endcsname{#7}%
	\addcontentsline{toc}{#1}{%
		\ifnum #2>\c@secnumdepth\else
		\protect\numberline{\csname the#1\endcsname}\fi #7}%
	\else\def\@svsechd{#6{\hskip #3\relax
			\@svsec #8\ifnum#2=2.\fi}%
		\csname #1mark\endcsname{#7}%
		\addcontentsline{toc}{#1}{%
			\ifnum #2>\c@secnumdepth \else
			\protect\numberline{\csname the#1\endcsname}\fi #7}}%
	\fi\@xsect{#5}}

\renewcommand\section{\@startsection {section}{1}{\z@}%
	{-10pt \@plus -1ex \@minus -.2ex}{.5ex }{\normalfont\Large\bfseries\sectionfont}}
\renewcommand\subsection{\@startsection{subsection}{2}{\z@}%
	{10pt\@plus 1ex \@minus .2ex}{-0.5ex \@plus .2ex}{\normalfont\large\bfseries\subsectionfont}}
\def\frontmatter@title@format{\titlefont\centering}%
\def\frontmatter@title@below{\addvspace{-5pt}}%

\renewcommand\NAT@biblabelnum[1]{#1.}
\renewcommand\NAT@citesuper[3]{\ifNAT@swa
	\unskip\hspace{1\p@}\textsuperscript{(#1)}%
	\if\relax#3\relax\else\ (#3)\fi\else (#1)\fi\endgroup}

\newcommand*\bib@heading{%
	\section{\refname}
	\fontsize{8}{10}\selectfont
}
\newcommand*\@openbib@code{%
	\advance\leftmargin\bibindent
	\itemindent -\bibindent
	\listparindent \itemindent
	\parsep \z@
}%
\newdimen\bibindent
\usepackage[usenames,dvipsnames,svgnaes,table]{xcolor}
\definecolor{col1}{rgb}{0.0, 0.30, 1.0}
\definecolor{col2}{rgb}{0.9, 0.0, 0.30}

\makeatletter
\begin{document}
\title{Thermoelectric Alchemy: Designing A Chemical Analog to PbTe with Intrinsic High Band Degeneracy and Low Lattice Thermal Conductivity}
\author{Jiangang He}
\altaffiliation{Contributed equally to this work}
\affiliation{Department of Materials Science and Engineering, Northwestern University, Evanston, IL 60208, USA}
\author{Yi Xia}
\altaffiliation{Contributed equally to this work}
\affiliation{Center for Nanoscale Materials, Argonne National Laboratory, 9700 South Cass Avenue, Lemont, Illinois 60439, United States} 
\author{S. Shahab Naghavi}
\affiliation{Department of Physical and Computational Chemistry, Shahid Beheshti University, G.C., Evin, 1983969411 Tehran, Iran}
\author{Vidvuds Ozoli\c{n}\v{s}}
\affiliation{Department of Applied Physics, Yale University, New Haven, CT, 06520, USA}
\affiliation{Yale Energy Sciences Institute, West Haven, CT, 06516, USA}
\author{Chris  Wolverton}
\email{c-wolverton@northwestern.edu}
\affiliation{Department of Materials Science and Engineering, Northwestern University, Evanston, IL 60208, USA}
\date{\today}

\begin{abstract}
Improving the figure of merit $zT$ of thermoelectric materials requires
simultaneously a high  power factor and low thermal conductivity. An effective
approach for increasing the power factor is to align the band extremum and achieve high band degeneracy ($\geq$ 12)
near the  Fermi level as realized in PbTe [\textcolor{blue}{Pei et. al. \textit{Nature} 473, 66 (2010)}], which  usually  relies on band
structure engineering, e.g., chemical doping and strain. However, very few materials could achieve such a high
band degeneracy without heavy doping or suffering impractical strain. By employing state-of-the-art first-principles methods with direct computation of phonon and carrier lifetime, we demonstrate that two new full-Heusler compounds Li$_2$TlBi and Li$_2$InBi, possessing 
a PbTe-like electronic structure, show
exceptionally high power factors ($\sim$ 20 mWm$^{-1}$K$^{-2}$ at 300 K) and low lattice thermal conductivities (2.36 and 1.55 Wm$^{-1}$K$^{-1}$) at room temperature. The Tl$^{+}$Bi$^{3-}$ (In$^{+}$Bi$^{3-}$) sublattice forms a rock-salt structure, and the additional two valence electrons from Li atoms essentially make these compounds isovalent with Pb$^{2+}$Te$^{2-}$.
The larger rock-salt sublattice of TlBi (InBi) shifts the valence band maximum from L point to the middle of the $\Sigma$ line, increasing the band degeneracy from fourfold to twelvefold. On the other hand, resonance bond in the PbTe-like sublattice and soft Tl-Bi (In-Bi) bonding interaction is responsible for intrinsic low lattice thermal conductivities.
Our results present a novel strategy of designing high performance thermoelectric materials.

\end{abstract}
\maketitle

\lettrine[lines=3, findent=3pt, nindent=0pt]{\color{BurntOrange}T}{}hermoelectric (TE) materials have important applications in energy harvesting, thermoelectric coolers, and thermal detectors as they
can directly convert heat into electricity and vise versa. High efficient TE materials are required
for practical applications and are characterized
by the figure of merit $zT=(S^2\sigma{T})/(\kappa_{\rm L}+\kappa_{\rm e})$,
where  $S$,   $\sigma$,  $\kappa_{\rm e}$,  $\kappa_{\rm L}$,   and  $T$  are   the  Seebeck
coefficient,  electrical conductivity,  electronic  thermal conductivity, lattice
thermal conductivity, and temperature, respectively. In order to maximize $zT$, both electronic transport properties and lattice
thermal conductivity have to be optimized carefully. Many strategies have been successfully used for the suppression of
$\kappa_{\rm L}$~\cite{toberer2011phonon}. However, there are fewer approaches that can effectively improve the electronic
properties, i.e., the power factor (PF=$S^2\sigma$)~\cite{pei2012band,he2017bi2pdo4,usui2017enhanced}. One effective route is to increase the band degeneracy ($N_{\rm v}$) and decrease the inertial effective mass ($m_{\rm I}^*$) simultaneously
since the figure of merit $zT$ of a material is proportional to $\frac{N_{\rm v}}{m_{\rm I}^*}$~\cite{pei2012band,zeier2016thinking}.
Although a  high  density  of  states  (DOS)
effective mass ($m_{\rm d}^*$ = $N_{\rm v}^{2/3}m_{\rm b}^*$) is preferred for generating a high $S$~\cite{snyder2008complex,heremans200627},
the band  effective mass $m_{\rm b}^*$ is  also concomitantly high in a material with low  $N_{\rm v}$, leading to a low electrical conductivity as $\sigma \propto \frac{\tau}{m_{\rm b}^*}$  ($\tau$ is the carrier lifetime)~\cite{zeier2016thinking}.

\begin{figure}
	\centering
	\includegraphics[width=1.0\linewidth]{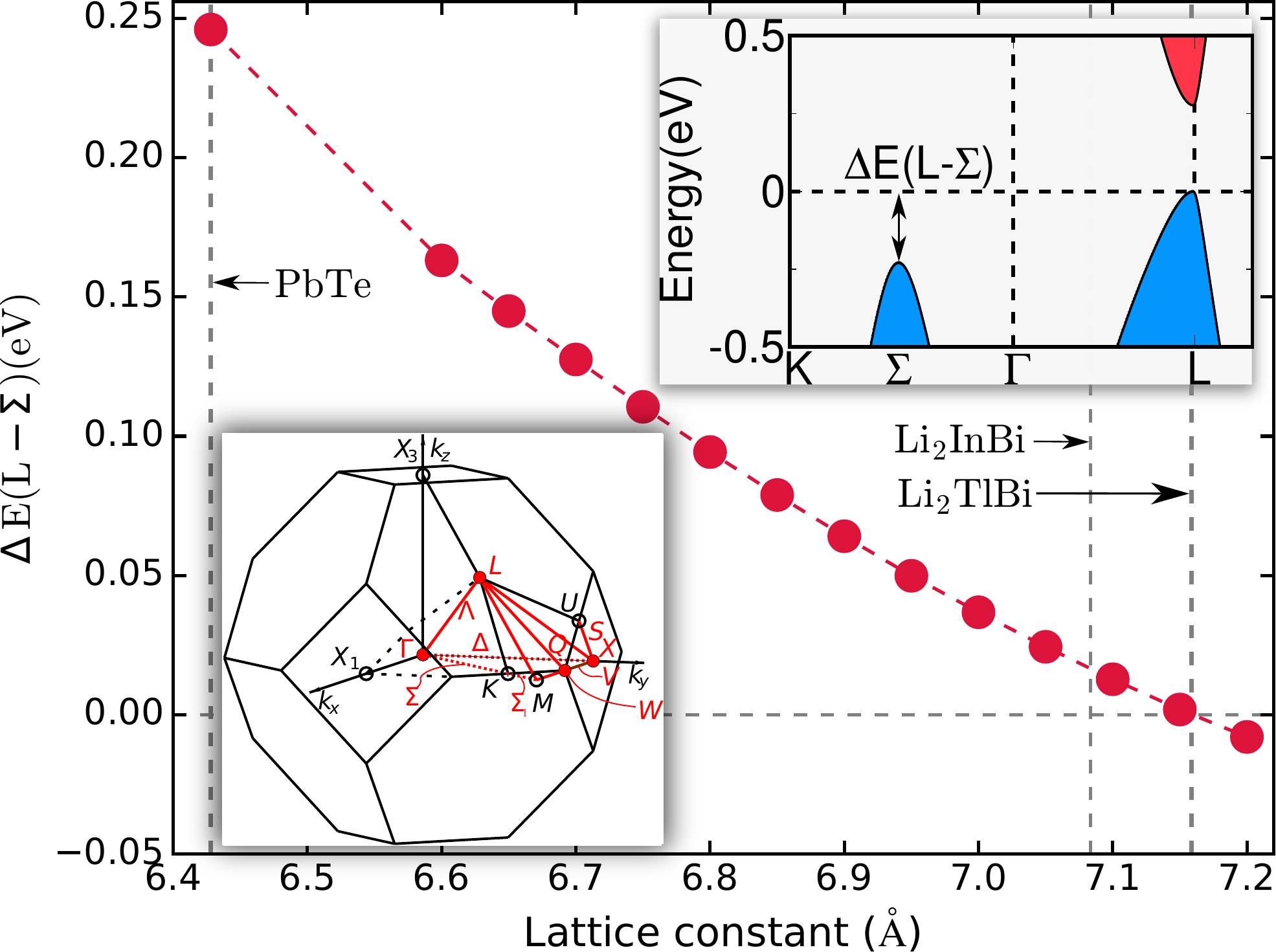}
	\caption{The valence band energy difference between L and $\Sigma$ of PbTe as a function of lattice constant. The vertical lines indicates the lattice constants of experimental PbTe and fully relaxed Li$_2$TlBi and Li$_2$InBi. Insets are the band structure (blue: valence band; red: conduction band) of PbTe along K-$\Gamma$-L direction and the symmetry pints/line of the first Brillouin zone of Fm$\bar{3}$m.}
	\label{PbTeband}
\end{figure}

A high value of $N_{\rm v}$ can be achieved either from a high valley multiplicity (the number of the carrier pockets of a band in the Brillouin zone) or a high orbital degeneracy (the number of bands with the same energy). Take the well studied TE material PbTe as an
example, once the second maximum of the valence band (the middle of the $\Sigma$ line, multiplicity is 12) is converged with the valence band maximum (at the L point, multiplicity is 4) by alloying an appropriate amount PbSe, a significant enhancement of $zT$ from 0.8 to 1.8 can be reached~\cite{pei2011convergence}. Unfortunately, most intrinsic semiconductors have very low valley multiplicity and heavy doping is required to align band valleys around the Fermi level. A high valley multiplicity usually only appears in cubic crystal systems where the valence band maximum (VBM) or conduction band minimum (CBM) is located in a low symmetry point of the Brillouin zone, such as the $\Sigma$ line of the rock-salt structure.
In addition to alloying, the band convergence could, in principle, be achieved through strain engineering. The lattice constant plays an important role on the alignment of $\Sigma$ and L~\cite{pei2011convergence,zhu2014band}. Consisting with the previous calculation~\cite{zhu2014band}, we found the expansion of PbTe lattice constant results in a remarkable decrease of the energy difference between $\Sigma$ and L, as depicted in Figure~\ref{PbTeband}.
The band maximum in the middle of $\Sigma$ aligns with that at L point when the lattice constant of PbTe (6.462 \AA~\cite{dalven1969review}) extends to 7.15 \AA. However, PbTe could not be stabilized for a such large strain ($\sim$ 11\%) in practice. Therefore, a new material design strategy is demanded.

The full Heusler (chemical formula $X_2YZ$; space group Fm$\bar{3}$m) structure is a face centered cubic crystal structure with the inter-penetration of $X_2$ cubic and $YZ$ rock-salt sublattices. The embedded cubic sublattice extends the bond length between $Y$ and $Z$ atoms of the rock-salt sublattice. Therefore, full Heusler structure is an idea candidate of designing extended PbTe.

In this work, two stable full-Heusler (FH) compounds with PbTe-like electronic structure, Li$_2$TlBi and Li$_2$InBi, are discovered by employing a high throughput $ab~initio$ thermal dynamic screening~\cite{kirklin2016high,PhysRevLett.117.046602}. The crystal structure of FH Li$_2$TlBi (Li$_2$InBi) is the interpenetration of Li$_2$ cubic and TlBi (InBi) rock-salt sublattices. The electronic structure of [Li$^{+}$]$_2$[Tl$^{+}$Bi$^{3-}$] ([Li$^{+}$]$_2$[In$^{+}$Bi$^{3-}$]) is isoelectronic with PbTe (Pb$^{2+}$Te$^{2-}$) since the electron donated by Li is delocalized in the whole system. However, the bond length of Tl-Bi (In-Bi) is considerably extended in the full Heusler lattice. Consequently, both the VBM and CBM of these two compounds are located in the middle of the $\Sigma$ line, with band degeneracy of $N_{\rm v}$=12 in the intrinsic compounds due to their large lattice constants ($\sim$ 7.15 \AA). Our transport calculations, including phonon-phonon and phonon-electron interactions, show that these two compounds have low $\kappa_{\rm L}$ and high PF at room temperature. 
Benefited from their low $\kappa_{\rm L}$ and high PF, Li$_2$InBi and Li$_2$TlBi are therefore identified as promising room temperature TE materials with $zT$ of 1.5 and 2.0 at 300 K, respectively.

\begin{figure}
	\centering
	\includegraphics[width=1.0\linewidth]{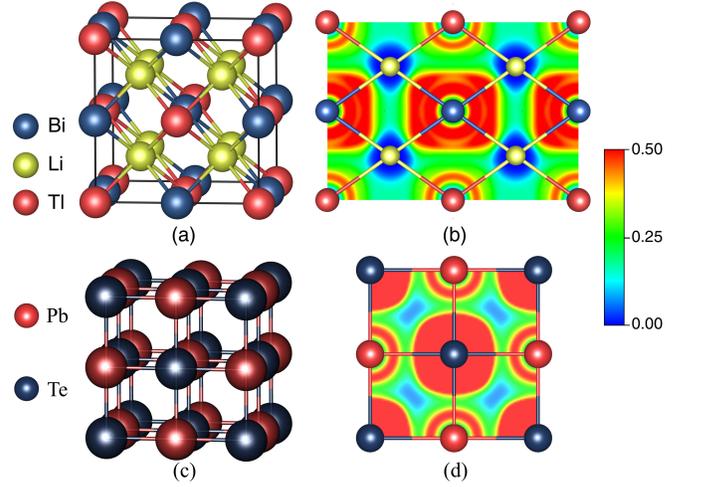}
	\caption{(a) crystal structure of full Heusler Li$_2$TlBi, (b) electron localization function (ELF) of Li$_2$TlBi in (110) plane, (c) crystal structure of PbTe, and electron localization function (ELF) of PbTe in (001) plane.}
	\label{structure}
\end{figure} 

\section{Results}
\subsection{Stability}
Our density functional theory (DFT) calculations show that Li$_2$TlBi is on the T = 0~K convex hull, which means it is thermodynamically stable, and Li$_2$InBi is just 2 meV/atom above the convex hull, which indicates it is thermodynamically weakly metastable, as constructed by the Open Quantum Material Database
(OQMD)~\cite{OQMD}. We also performed a ground state crystal structure search by using 21 distinct $X_2YZ$ prototype structures, and we find the FH structure is the lowest energy structure for both Li$_2$InBi and Li$_2$TlBi. Phonon calculations show that Li$_2$TlBi and Li$_2$InBi are dynamically stable at T = 0 K.
The formation free energy ($\Delta{G} = \Delta{H} - T\Delta{S}$) calculation including vibrational entropy difference ($\Delta{S}$) between Li$_2$InBi and its competing phases shows that Li$_2$InBi is thermodynamically stable above room temperature, see supplementary Figure \textcolor{blue}{1}.

\subsection{Electronic structure}
\begin{figure}
	\setlength{\unitlength}{1cm}
	\includegraphics[width=0.8\columnwidth]{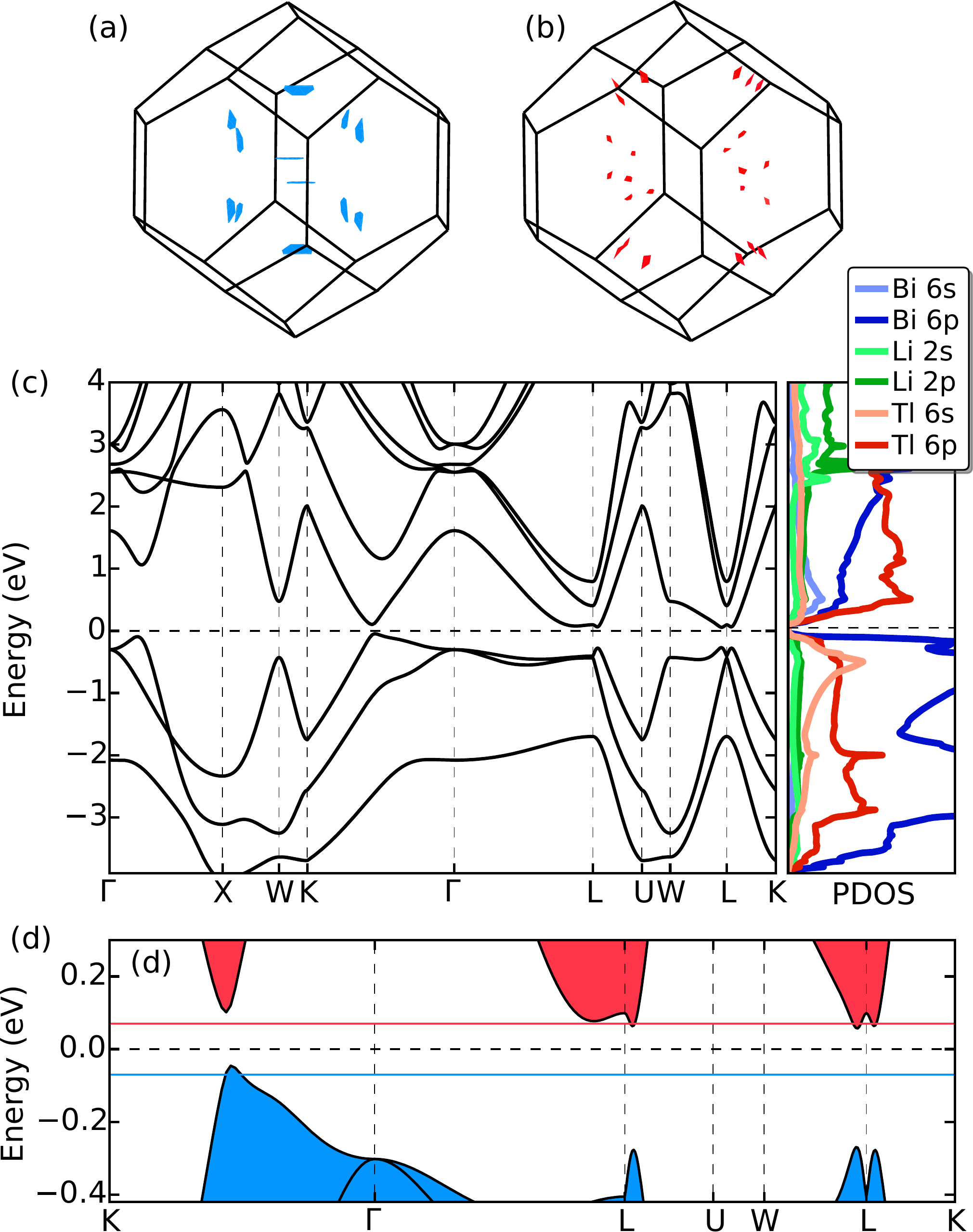}
	\caption{Fermi   surface   of   hole   (a)  and   electron   (b)   doped
		Li$_2$TlBi.  (c)  Band  structure  (left)   and  density  of  state  (right)  of
		Li$_2$TlBi. (d) Expanded view of the band structure aground the Fermi level. The size of Fermi pockets
		in (b) are re-scaled by factor of two for the purpose of visualization.}
	\label{bandstructure}
\end{figure}

The main band structure characters of Li$_2Y$Bi ($Y$=In and Tl) FH are determined by [$Y^{+}$Bi$^{3-}$]$^{2-}$, which are isoelectronic with Pb$^{2+}$Te$^{2-}$ even though In/Tl (Bi) is cubic-coordinated with eight Li atoms as the nearest neighbors and octahedrally-coordinated with six Bi (In/Tl) as the next nearest neighbor. This is because Li is the most electropositive element in these compounds and it provides its 2$s$ electron to the crystal system. As shown in Figure~\ref{bandstructure}, Li 2$s$ electron is completely delocalized in these compounds and therefore it has very limit influence on electronic structures of Li$_2Y$Bi ($Y$=In and Tl) except for raising the Fermi level and opening the band gap, which is similar to
the Li stabilized quaternary Heusler semiconductors~\cite{he2018designing}. We further verified this conclusion by performing band structure calculations for 
[TlBi]$^{2-}$ with the -2 charge balanced by a +2 Jellium background, see supplementary Figure~\textcolor{blue}{2}. Since Tl is the nearest neighbor of Pb and Bi is close to Te in the periodic table, Li$_2$InBi and Li$_2$TlBi have very similar band structures with PbTe, as shown in Figure~\ref{bandstructure}. Interestingly, Li$_2$InBi and Li$_2$TlBi have much larger lattice constants than PbTe because of the inserted Li$_2$ cubic sublattice, which plays an important role in raising the energy level of the VBM at the middle of $\Sigma$ line. As a consequence, the $N_{\rm v}$ of Li$_2Y$Bi reaches to 12,
as observed in the PbTe under significant hydrostatic expansion. At the same time, the large bonding distance (softer bonding interaction) between Bi and $Y$ ($Y$=In and Tl) contributes to reducing $\kappa_{\rm L}$ as we will see later~\cite{zeier2016thinking}.

%
Since Li$_2$InBi has a very similar electronic structure to Li$_2$TlBi, we only take Li$_2$TlBi as an example here.
The electronic structure of Li$_2$TlBi is shown in Figure~\ref{bandstructure} (the band structure of Li$_2$InBi is shown in supplementary Figure~\textcolor{blue}{3}).
Li$_2$TlBi is a small band gap semiconductor (PBE:~0.06\,eV;
HSE06:~0.18\,eV, including the spin orbit coupling (SOC). These calculated gaps are well comparable with many high $zT$ TE materials,
such as PbTe:~0.19 eV~\cite{PhysRevB.89.205203} and CoSb$_3$:~0.05~$\sim$~0.22
eV~\cite{PhysRevB.50.11235,PhysRevB.58.15620}). In Li$_2$TlBi, the band gap opens between the fully occupied Bi
6$p$ and fully unoccupied Tl 6$p$ states due to charge transfer from Tl to Bi. Tl atom loses
its one 6$p$ electron to the more electronegative Bi atom and becomes Tl$^{+}$, and
its 6$s^2$ electrons are deeply ($\sim$ -5\,eV below the Fermi level) buried
below the Bi 6$p$ orbitals (valence bands, from -4 to 0\,eV), forming
stereochemically inactive lone pair electrons. Two electropositive Li atoms lose their 2$s$ electrons to Bi as
well. Therefore, the 6$p$ orbitals of Bi$^{3-}$ (from -4 to 0\,eV below the Fermi
level) are fully filled with six electrons. The splitting of three
occupied Bi 6$p$ orbitals into two groups, $\sim$ -2\,eV (single degeneracy) and
$\sim$ -0.5\,eV (double degeneracy) at the $\Gamma$ point is due to SOC. The
conduction bands are mainly from the Tl$^{+}$ 6$p$ and Bi$^{3-}$ 6$p*$ orbitals. The electron
localization function (ELF) is shown in Figure~\ref{structure}. We can clearly
see that Bi and Li atoms have the highest and lowest ELF values, respectively,
consistent with our electronic structure analysis that they are behaving as Bi$^{3-}$
anions and Li$^{+}$ cations. Therefore, Li$_2$TlBi is an ionic compound,
similar to the half Heusler LiMgN~\cite{kandpal2006covalent}. The inactive lone-pair electrons
of Tl$^{+}$ 6$s^2$ are also clearly seen in Figure~\ref{structure}. We also can
clearly see that the ELF of Tl and Bi in Li$_2$TlBi are very similar to Pb and Te in PbTe.

\begin{figure}
	\setlength{\unitlength}{1cm}
	\includegraphics[width=1.0\columnwidth]{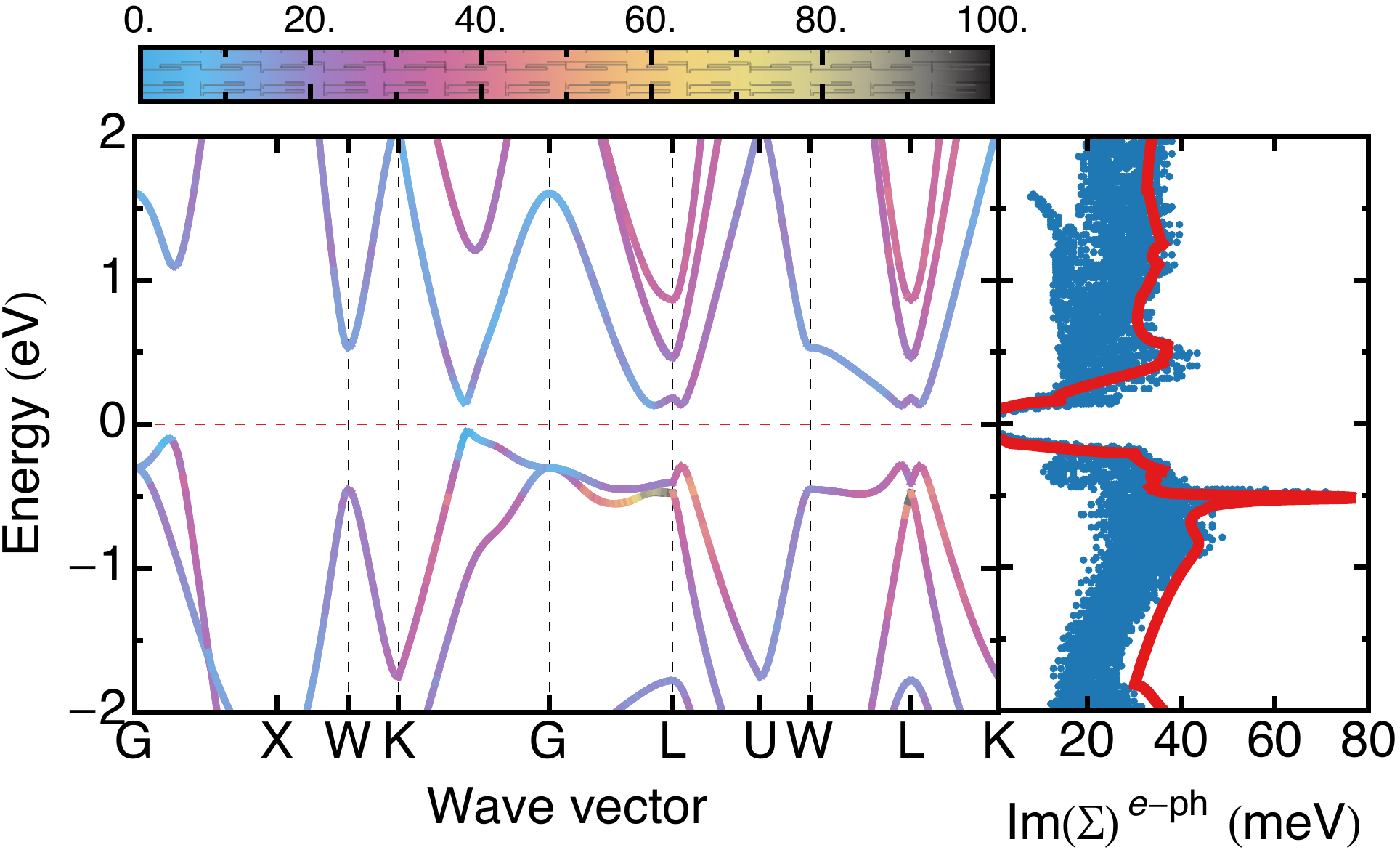}
	\caption{A heat map of the imaginary part of the electron-phonon self-energy (Im($\Sigma$)) of Li$_2$TlBi at 300 K (left) and mode-dependent Im($\Sigma$) compared with scaled density of states (DOS) (right).}
	\label{elph}
\end{figure}

As expected from the previous analysis, a remarkable character of Li$_2$TlBi band structure is the VBM is located
in the middle of $\Sigma$ line of the first Brillouin zone of the FCC full Heusler structure (Fm$\bar{3}$m),
which leads to an unexpected high valley degeneracy ($N_{\rm v}$ =12), see the Fermi surface in Figure~\ref{bandstructure}. Hence the $N_{\rm v}$=12 of the VBM reaches a record high value, which only has been previously matched in the heavily doped PbTe and
CoSb$_3$ systems~\cite{pei2011convergence,tang2015convergence}.
The second VBM, which is $\sim$ 40 meV lower than VBM, is located at the middle of the $\Delta$ line (between $\Gamma$ and X)
and possesses a valley degeneracy of 6. Therefore, an extremely high $N_{\rm v}$=18 is reachable in Li$_2$TlBi by means of hole doping.
Although the CBM is located at L with the valley degeneracy of 4,
the energy difference between CBM and the second CBM (in the middle of
$\Sigma$ line) is just 7 meV. Therefore the $N_{\rm v}$ of the conduction band
can potentially reach as high as 16 through light electron doping.
The Fermi surface of valence band and conduction bands are displayed in Figure~\ref{bandstructure}. As mentioned above, although the band effective masses ($m_{\rm b}^*$)
for the VBM and CBM are small, which imply high carrier mobilities as $\mu
\propto \frac{\tau}{m_{\rm b}^*}$, the Seebeck coefficient $S \propto m_{\rm d}^*$ still can be very high,
provided $N_{\rm v}$ is sufficiently high, since $m_{\rm d}^*$ is related to the band
effective mass by $m_{\rm d}^*=N_{\rm v}^{2/3}m_{\rm b}^*$.

\begin{figure}
	\includegraphics[width=1.0\linewidth,angle=0]{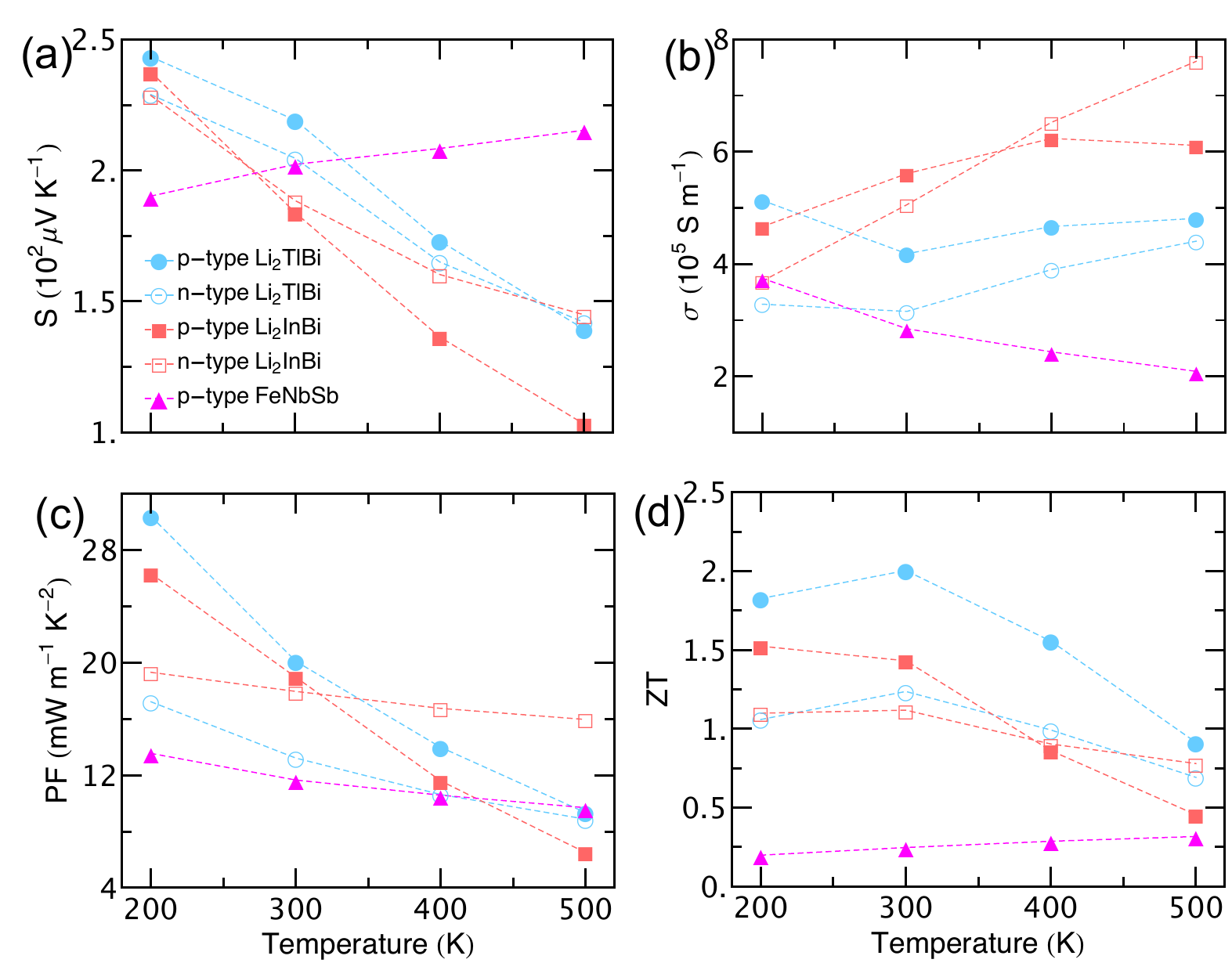}
	\caption{The Seebeck coefficient ($S$), electrical conductivity ($\sigma$), electron  thermal conductivity ($\kappa_{\rm e}$), Power  factor  (PF, $\sigma{S^2}$), and $zT$  of Li$_2$TlBi and Li$_2$InBi at carrier concentrations that give rise to maximum $zT$ at 200, 300, 400 and 500 K, comparing with $p$-type FeNbSb. The lattice thermal conductivity of FeNbSb used to compute $zT$ is extracted from Ref.~\cite{He29112016}.}
	\label{ZTelph}
\end{figure}

\subsection{Electron transport}
To quantitatively characterize the electron transport of Li$_2Y$Bi ($Y$=In and Tl), we calculate the $S$ and $\sigma$ based on the Boltzmann transport equation under relaxation time approximation. We assume that the predominant carrier scattering mechanisms at 200 K and above are all based on phonons: (1) deformation potentials of acoustic and optical phonons and (2) Fr\"ohlich coupling due to polar optical phonons~\cite{Freik_2002,Vineis_2008}. Since the best thermoelectric efficiency is always achieved in the heavily doped region where the scattering on polar optical phonons are sufficiently screened and the dielectric constant is usually large in narrow band gap semiconductors~\cite{Freik_2002,Vineis_2008,Leite_2013}, we mainly take into account deformation potential scattering based on first-principles calculated electron-phonon interaction (EPI) matrix elements.
As shown in Figure~\ref{elph} for the representative compound Li$_{2}$TlBi, the imaginary part of the electron self-energy Im($\Sigma$) shows a strong energy dependence and is roughly proportional to the density of electronic states. States with a long lifetime appear near the VBM and CBM. This indicates the lifetime is linked to the phase space availability for electronic transitions, that is, electrons and holes near band edges are less scattered due to limited phase space~\cite{Giustino_2007}.

To validate our calculations, we also computed the thermoelectric properties for a well studied $p$-type half-Heusler (HH) compound FeNbSb, for which a PF as large as 10.6 mWm$^{-1}$K$^{-2}$ was recently measured at room temperature~\cite{He29112016}. Figure~\ref{ZTelph} (c) shows that our calculation considering electron-phonon coupling predicts a maximum PF of 11.7 mWm$^{-1}$K$^{-2}$ for FeNbSb at 300 K, representing the upper limit without considering other scattering sources such as defects and grain boundaries. The good agreement between our calculation and the experiment confirms our assumption that electron-phonon coupling dominates carrier scattering in this system. It is noteworthy that the optimal PF of FeNbSb is significantly higher than that of PbTe at 300 K \cite{Pei1400486,Pei:2012aa,pei2011convergence}.

Next, we illustrate the ultrahigh PFs of Li$_{2}$TlBi and Li$_{2}$InBi by comparing to FeNbSb. Despite that $S$ is generally much higher in FeNbSb due to its larger band gap of 0.54 eV compared to 0.18 eV (Li$_{2}$TlBi) and 0.15 eV (Li$_{2}$InBi), the $S$ of Li$_{2}$TlBi and Li$_{2}$InBi is comparable with FeNbSb at optimal carrier concentration that leads to maximum $zT$, particularly at 300 K, as shown in Figure~\ref{ZTelph}(a). The strong bipolar effect further suppresses $S$ of Li$_{2}$TlBi and Li$_{2}$InBi at higher temperatures. However, owing to the smaller band effective mass and high valley degeneracy ($N_{\rm v}$), both Li$_{2}$TlBi and Li$_{2}$InBi have significantly higher $\sigma$ than FeNbSb from 300 to 500 K with a carrier concentration about one order of magnitudue lower than FeNbSb (see supplementary Figure~\textcolor{blue}{4} and \textcolor{blue}{5}). As a consequence, Li$_{2}$TlBi (Li$_{2}$InBi) achieves exceptional PFs of 30.4/20.1 (26.3/19.0) mWm$^{-1}$K$^{-2}$ at 200/300 K, nearly twice that of FeNbSb at 300 K. The outperformance of Li$_{2}$TlBi and Li$_{2}$InBi over FeNbSb is due to a comparable $S$ and a higher $\sigma$ at the optimized carrier concentrations, supporting our previous discussion.

\subsection{Lattice thermal conductivity}
\begin{figure}
	\setlength{\unitlength}{1cm}
	\includegraphics[width=1.0\columnwidth]{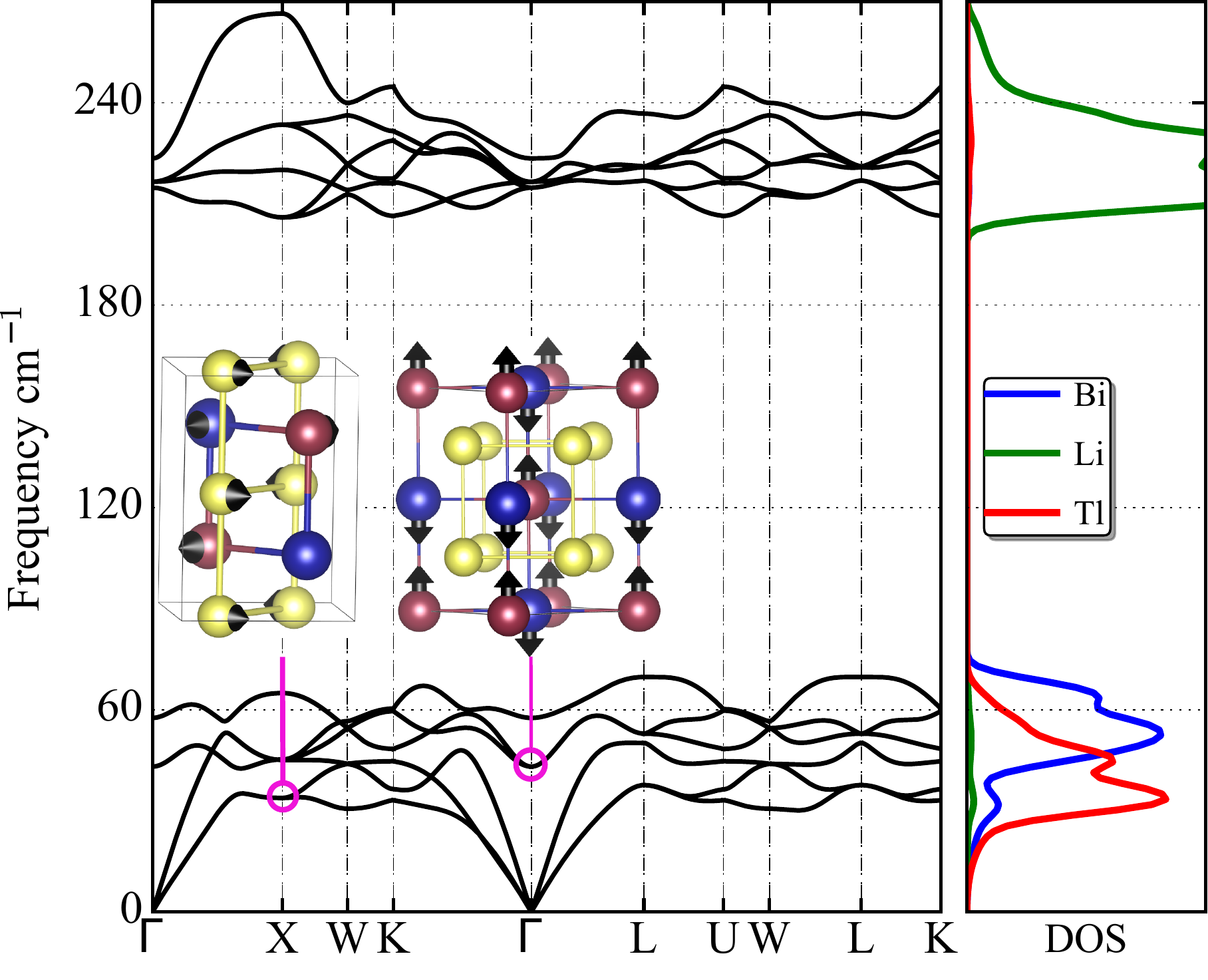}
	\caption{Phonon dispersion (left) and phonon density of state (right) of
		Li$_2$TlBi. The longitudinal optical (LO)  and transverse optical (TO) splitting
		is included. Inset is the first Brillouin zone of Fm$\bar{3}$m.}
	\label{phonon}
\end{figure}

The Li$_2$TlBi (Li$_2$InBi) primitive cell contains  4 atoms  and  therefore  12  phonon branches.   The  mode
decomposition in the zone center ($\Gamma$ point) is 3T$_{1u}$~$\oplus$~1T$_{2g}$.
As  shown in  Figure~\ref{phonon}, the low-frequency  phonon modes
are mainly from the stereochemically inert lone-pair Tl$^{+}$ cation instead of the heaviest atom Bi, which is
consistent  with the  weaker  bonding  between Tl  atom  and  its neighbors.
As expected, the light  lithium atom has much higher phonon frequencies 200 $\sim$ 250
cm$^{-1}$ and its phonon bands are completely separated from Tl and Bi. These
compounds possess two main differences from the previously reported alkali
metal based rattling ($R$) Heusler~\cite{PhysRevLett.117.046602}: i) higher acoustic phonon
frequencies, and ii) higher frequency of crossing bands between acoustic and
optical modes, meaning Tl (In) atom has a slightly stronger interaction  with its
neighbors than $R$ Heusler compounds.

\begin{figure}
	\setlength{\unitlength}{1cm}
	\includegraphics[width=1.0\columnwidth,angle=0]{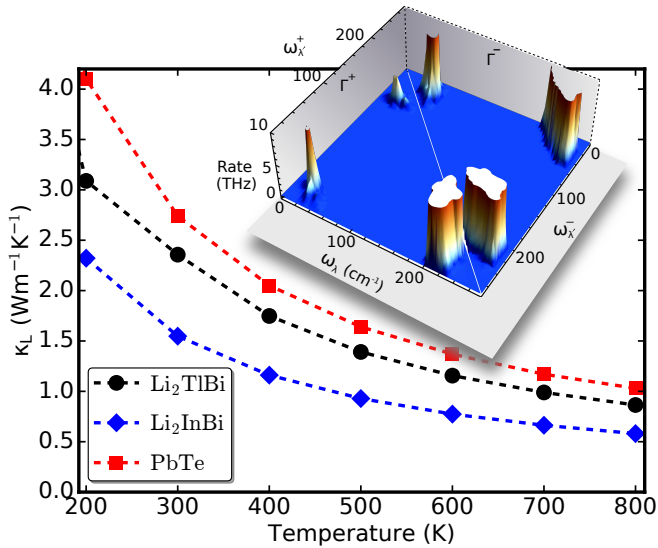}	
	\caption{Lattice  thermal  conductivity  of Li$_2$TlBi  as  function  of
		temperature. Insert  is the phonon  scattering rates in  absorption ($\Gamma^+$:
		$\lambda + \lambda' \rightarrow \lambda''$) and emission ($\Gamma^-$: $\lambda''
		\rightarrow \lambda + \lambda'$) processes.}
	\label{Kappa}
\end{figure}

The lattice thermal conductivity $\kappa_{\rm L}$ is calculated by using first-principles compressive sensing lattice dynamics
(CSLD) and solving the linear Boltzmann equation (see~\textcolor{blue}{Methods} for details) and the results are shown in Figure~\ref{Kappa}.
Owing to the cubic symmetry, $\kappa_{\rm L}$ of Li$_2$TlBi and Li$_2$InBi are isotropic
($\kappa_{\rm L}^{xx}$=$\kappa_{\rm L}^{yy}$=$\kappa_{\rm L}^{zz}$=$\kappa_{\rm L}$) and the calculated
$\kappa_{\rm L}$ are 2.36 (1.55) Wm$^{-1}$K$^{-1}$ at 300~K and 0.55 (0.52) Wm$^{-1}$K$^{-1}$ at
900~K for Li$_2$TlBi (Li$_2$InBi), which are much lower than most FH and HH ($\geq$ 7 Wm$^{-1}$K$^{-1}$~\cite{Chen2013387})
compounds without doping or nanostructuring and also lower
than PbTe (2.74 at 300 K and 0.91 Wm$^{-1}$K$^{-1}$ at 900 K at the same computational level). 

Similar to PbTe, Li$_2$InBi and Li$_2$TlBi have low-lying transverse optical modes (TO), implying the resonant bonding~\cite{lee2014resonant},
as expected from the similarity of electronic structures. The long-range interaction caused by the resonant bonding leads to
strong anharmonic scattering and large phase space for three-phonon scattering processes and, therefore, significantly suppresses
thermal transport~\cite{lee2014resonant}. Moreover, the weak Tl-Bi (In-Bi) bonding resulting from the large bonding distance between Tl and Bi (In and Bi) gives rise to
low group velocities. Finally, the high-frequency optical modes associated with the Li atoms provide extra scattering channels for low-lying acoustic modes.

The mechanism of the strong scattering of heat carrying
acoustic phonon  modes can be directly understood from phonon-phonon interactions. We  show  the  phonon-phonon  scattering rates  in  the
absorption ($\Gamma^+$: $\lambda + \lambda' \rightarrow \lambda''$) and emission
($\Gamma^-$:   $\lambda''  \rightarrow   \lambda  +   \lambda'$)  processes   in
Figure~\ref{Kappa}. The  low-frequency acoustic phonons are  mainly scattered by
the low-frequency  optical modes in the absorption  process, while the optical
modes decompose largely into low  energy acoustic modes in the emission  process. This scattering picture is
similar to the alkali metal based $R$-Heusler compounds~\cite{PhysRevLett.117.046602}.

\section{Discussion}
Using our calculated $\kappa_{\rm L}$, $S$, $\sigma$, and $\kappa_{\rm e}$ within DFT framework by explicitly including phonon-electron
and phonon-phonon interactions, the maximum figure of merit $zT$ of Li$_{2}$TlBi and Li$_{2}$InBi
are estimated to be 2.0 and 1.4 at 300 K for hole doping ($p$-type), respectively, which implies that
Li$_{2}$TlBi is the TE material with the highest $zT$ at room temperature.
The optimized carrier concentrations for the maximum $zT$ at 300 K are 1.3$\times$10$^{-19}$ and 1.6$\times$10$^{-19}$ cm$^{-3}$ for Li$_{2}$TlBi and Li$_{2}$InBi, respectively, which is similar to PbTe \cite{Pei1400486,Pei:2012aa} but one order of magnitude lower than FeNbSb~\cite{He29112016} and is much easier to achieve.
Our calculated $zT$ might be underestimated by excluding the
phonon-phonon interaction beyond the third-order and phonon scattering by defects in our calculations.
Furthermore, the $zT$ of these full Heusler materials could be even further enhanced by suppressing
heat transport through nano-structuring precipitates and grain boundaries as achieved in other Heusler compounds.
Owing to the small band gap, however, the maximum $zT$ of Li$_{2}$TlBi and Li$_{2}$InBi are at room temperature, see Figure~\ref{ZTelph}.
The drop down of the $zT$ at higher temperature is mainly due to the decreased PF by the bipolar effect.  
We also note that the electron doped ($n$-type) Li$_{2}$TlBi and Li$_{2}$InBi have high $zT$ at room temperature as well, due to the high conduction band degeneracy (at $\Sigma$ line and L point) and low lattice therm conductivity. The material with high $zT$ for both hole and electron doping is very important for fabricating TE devices. Therefore, Li$_{2}$TlBi and Li$_{2}$InBi are very promising materials of TE device operating at room temperature.

\vspace{1.0cm}
In summary, we discovery two promising room-temperature TE materials, Li$_2$TlBi and Li$_2$InBi Heuslers, by creating the analogs with isovalent electronic structures to PbTe but much expanded lattices. We demonstrate Li$_2$TlBi and Li$_2$InBi possess intrinsic high PFs and low $\kappa_{\rm L}$ by using the electron
Boltzmann transport theory with {\sl ab~initio} carrier relaxation time from electron-phonon coupling and phonon transport theory with phonon lifetime from first-principles compressive sensing lattice
dynamics. The high $zT$ of the $p$-type Li$_2$TlBi ($\sim$ 2.0) and Li$_2$InBi ($\sim$ 1.4) at room temperature are mainly due to the extremely high $N_{\rm v}$ and the low $\kappa_{\rm L}$ caused by the resonant bonding as observed in PbTe and weak bonding interactions, respectively. Our results not only 
present two high $zT$ room-temperature TE materials and highlight the importance of band degeneracy in enhancing PF, but also provide a novel routine for designing high-performance TE materials.

\section{Methods}
In this study, most DFT calculations are performed using the Vienna {\sl ab initio} Simulation Package (VASP)~\cite{vasp1,vasp2}. The projector augmented wave (PAW~\cite{PAW-Blochl-1994,VASP-Kresse-1999}) pseudo potential, plane wave basis set, and Perdew-Burke-Ernzerhof (PBE~\cite{PBE}) exchange-correlation functional were used. The qmpy~\cite{OQMD} framework and the Open Quantum Material Database~\cite{OQMD} was used for convex hull construction. The lowest energy structure of Li$_2Y$Bi were confirmed by prototype structure screening~\cite{PhysRevLett.117.046602}. The lattice dynamic stability was investigated by performing frozen phonon calculation as implemented in phonopy package~\cite{phonopy}. The band gap was computed by means of the screened hybrid functional HSE06~\cite{heyd2003hybrid}, including spin orbit coupling (SOC).
The compressive sensing lattice dynamics~\cite{PhysRevLett.113.185501} technique was employed to obtain the third-order force constants, which were used to
iteratively solve the linearized phonon Boltzmann equation
with the ShengBTE package~\cite{ShengBTE_2014}.
The carrier lifetime due to electron phonon coupling was computed by using Quantum Espresso and Electron-phonon Wannier (EPW) codes with SOC included ~\cite{giannozzi2009quantum,ponce2016epw}. Thermoelectric properties were computed using BoltzTrap~\cite{Madsen:2006aa} with adjusted band gap from HSE06 calculations.


\section{Acknowledgment}
J.~H. and C.~W. (stabilities and electronic structures calculations) acknowledge support by the U.S. Department of Energy, Office of Science and Office of Basic Energy Sciences, under Award No. DE-SC0014520. Y.~X. (lattice thermal conductivity and electron transport calculations), S.S.~N. (electronic structure analysis), and V.~O. (electronic structure analysis) were supported by US Department of Energy, Office of Science, Basic Energy Sciences, under grant DEFG02-07ER46433. 
This research used resources of the National Energy Research Scientific Computing Center, a DOE Office of Science User Facility supported by the Office of Science of the U.S. Department of Energy under Contract No. DE-AC02-05CH11231.



\section{Author contributions}
The research was conceived and designed by J.H., V.O., and C.W.
High throughput DFT screening, stabilities, and electronic structure calculations
were carried out by J.H. Thermoelectric properties calculations were conducted by Y.X.
Analysis of the data was performed by J.H., S.S.N., and Y.X.
All authors discussed the results contributed to writing the manuscript.

\vspace{0.5cm}
\section{Additional information}
Supplementary information is available in the online version of the paper.

\section{Competing financial interests}
The authors declare no competing financial interests.

\bibliographystyle{aipnum4-1}   
\bibliography{LTB,electron_phonon_interaction,pbte_general}

\end{document}